# PDS: Deduce Elder Privacy from Smart Homes

Ming-Chang Lee[1], Jia-Chun Lin[1,*], and Olaf Owe[2]

*Department of Information Security and Communication Technology, Norwegian University of Science and Technology,
Ametyst-bygget, 2815 Gjøvik, Norway
E-mail: ming-chang.lee@ntnu.no
E-mail: jia-chun.lin@ntnu.no*

[2] *Department of Informatics, University of Oslo,
Gaustadallèen 23 B
N-0373 Oslo, Norway
E-mail: olaf@ifi.uio.no*

21st January 2020

# PDS: Deduce Elder Privacy from Smart Homes

Ming-Chang Lee[1], Jia-Chun Lin[1,*], and Olaf Owe[2]

[1] *Department of Information Security and Communication Technology, Norwegian University of Science and Technology,*
*Ametyst-bygget, 2815 Gjøvik, Norway*
*E-mail: ming-chang.lee@ntnu.no*
*E-mail: jia-chun.lin@ntnu.no*

[2] *Department of Informatics, University of Oslo,*
*Gaustadalléen 23 B*
*N-0373 Oslo, Norway*
*E-mail: olaf@ifi.uio.no*

**Abstract**

With the development of IoT technologies in the past few years, a wide range of smart devices are deployed in a variety of environments aiming to improve the quality of human life in a cost efficient way. Due to the increasingly serious aging problem around the world, smart homes for elder healthcare have become an important IoT-based application, which not only enables elders' health to be properly monitored and taken care of, but also allows them to live more comfortably and independently in their houses. However, elders' privacy might be disclosed from smart homes due to non-fully protected network communication. To show that elders' privacy could be substantially exposed, in this paper we develop a Privacy Deduction Scheme (PDS for short) by eavesdropping sensor traffic from a smart home to identify elders' movement activities and speculating sensor locations in the smart home based on a series of deductions from the viewpoint of an attacker. The experimental results based on sensor datasets from real smart homes demonstrate the effectiveness of PDS in deducing and disclosing elders' privacy, which might be maliciously exploited by attackers to endanger elders and their properties.

## 1. Introduction

In recent years, the advances in the semiconductor technology enables to cost-efficiently integrate wireless network connectivity in embedded processors and sensors, which leads to the development of Internet of Things (IoT) [1]. In IoT paradigm, many different objects or devices that surround us connect together using Radio Frequency IDentification (RFID) and sensor network technologies (e.g., Z-Wave and ZigBee). IoT connects more and more devices every day. It is expected that we will have 24 billion IoT devices by 2020 [2]. IoT enables different smart environments with different smart devices, including sensors, actuators, displays, and computational components to monitor and control the environment, and interact with users to provide them with automated, customized, and comfortable services [3][4]. For instance, a smart environment could be a smart home, smart community, smart building, smart city, or smart grid. IoT also enables numerous applications, such as healthcare [4][5][6], energy conservation [7], home automation [8], remote access services [3][9], agriculture [10][11], security [12][13][14], surveillance [9][15][16][17], and transportation [18].

Among these applications, elder healthcare attracts a lot of attentions in recent years since aging population becomes a serious problem [4][19]. According to the Population Reference Bureau report [20], the number of Americans ages 65 and older is 46 millions today and will increase to over 98 millions by 2060. The UN Department of Economic and Social Affairs Population Division also predicts that there will be more elderly people than children, particularly in Europe. The ratio 15-year-olds to 65-year-olds will be 4:1 in 2050 [21]. Using smart home technology, elders' health can be properly taken care of, and the corresponding healthcare cost can be dramatically reduced. It also allows elderly people to live more comfortably and independently in their houses and prevent them from social isolation.

In most cases, a smart home for elder healthcare consists of a smart hub, motion sensors, door sensors, and other sensors, which all connect together via Z-Wave or ZigBee (see the left part of Figure 1). The hub also connects to a router via WiFi or Ethernet so as to communicate with a cloud server on Internet. These sensors provide rich information about the house environment and its residents, i.e., elders. The data sensed by different sensors is transmitted via Internet to the cloud server and analyzed. When an accident happens, e.g., an elder passes out on the floor, subsequent actions will be performed to timely deal with the accident. However, during the data transmission, elders' privacy such as their activities of daily living (ADLs), which refer to basic self-care activities for determining the independence of elderly people in their daily lives [22], might be exposed if attackers employ sophisticated eavesdropping tools [23][24][25][26]. Researches [23][24][27] have shown that many IoT devices and networks might disclose homeowners' privacy, including what TV programs they are watching, what sensors have been triggered, where their homes might be, and if they are home or not.

To show that elders' privacy could be seriously disclosed from smart homes, we extend PMA (Privacy Mining Approach for short) [47] and develop in this paper a Privacy Deduction Scheme (PDS for short) from the viewpoint of an attacker. PDS identifies elders' activities in smart homes from sensor reading, particularly their movements inside the houses. Based on identified activities, PDS infers how sensors spatially related to each other and then derives a global sensor topology, which reveals how sensors are deployed in the smart home. By conducting a series of deductions based on data mining, PDS is able to infer sensor locations and elders' daily routines without physical invasion.

To demonstrate the deduction performance of PDS, we apply PDS on two real smart homes with different layouts and different combinations of residents/elders. The experimental results show that, no matter in which case, PDS is able to deduce a global sensor topology that corresponds to the real one, infer most sensors in bedrooms, kitchen/dining room, and entrances with a high accuracy rate, and obtain the elders' daily routines in these places.

The rest of this paper is organized as follows. Section 2 describes the related work. Section 3 presents the methodology of PDS. In Section 4, extensive experiments are conducted and experimental results are discussed. Sections 5 and 6 discuss and conclude this paper, respectively.

## 2. Related Work

During the last few years, many security and privacy problems have arisen across IoT devices and networks, from networked light bulbs that can provide back doors into WiFi networks, to baby monitors that allow hackers to easily access the livestreams, and to network refrigerators that have been used in DDoS attacks [27]. Researchers investigated commercial off-the-shelf (COTS) IoT devices and found that these devices might disclose homeowners' privacy. As an example, Yoshigoe et al. [24] studied Samsung SmartThings platform and discovered that the network traffic sent from SmartThing Hub significantly reduces when homeowners are not at home. Such information not only reveals homeowners' privacy, but also might endanger homeowners and their properties. The authors also found that each kind of sensor (e.g., door sensors, light sensors, and motion sensors) has a unique traffic (handshaking) pattern between the SmartThings Hub and a cloud server. Therefore, even though the traffic is encrypted by SSL, attackers are still able to guess the corresponding sensor types and know which sensors are triggered. Furthermore, homeowners' location details might be revealed by the IP addresses of their smart devices [23], and sensitive information about smart homes and the residents might be leaked due to insecure communication [28]. In this paper, we further show that it is possible to disclose even more privacy by conducting a series of deductions on sensor reading from smart homes.

Many researchers proposed activity recognition methods to recognize common human activities from sensor reading (i.e., a low-level sensor dataset). Based on different requirements of smart environments, activity recognition can be either vision-based or sensor-based [29]. The former uses visual sensing facilities, such as video cameras, to monitor people's behavior and environmental changes. This approach is well known for suffering from privacy and ethics issues [6] [29][30]. The latter uses embedded sensors (such as motion sensors and door sensors) or wearable sensors (such as RFID tags attached to homeowners). Our focus in this paper is to expose elders' privacy from sensor data sent from embedded sensors since we are interested in understanding how elders move from sensors to sensors, which allows us to deduce the global sensor deployment of the corresponding smart homes.

Activity recognition can be divided into two major categories: supervised learning and unsupervised learning. In a supervised activity recognition method [31][32][33], each sensor event has its activity label, such as personal hygiene, enter home, bathing, bed to toilet, etc. For examples, Zhao et al. [34] proposed conditional random fields (CRFs)-based classifier to recognize human activities, and Inomata et al. [35] utilized Dynamic Bayesian Network (DBN) framework to recognize activities from interaction data collected by a RFID tag system. The authors in [36] used static Bayesian model and DBN, both with *k*-Observation history matrix, to recognize activities of ADLs from sensor readings, where $k$ is the number of timesteps they look into the past. Lu et al. [6] proposed a method that extracts latent features from sensor data by using Beta Process Hidden Markov Model (BP-HMM). However, it is well known that all the supervised methods suffer from a problem of manually labeling for all activities in training phase, which is time-consuming and laborious [19].

To solve this problem, unsupervised methods [5][19][37][38][39] have been proposed to automatically recognize human activities in smart homes. For instances, Rashidi et al. [5] presented an unsupervised method for discovering and tracking activities that homeowners normally perform as part of their daily routines in smart homes. Gu et al. [38] proposed a fingerprint-based algorithm to recognize activities such that it can mine large number of activity models on the web without manual labeling. The authors in [37] introduced an Emerging Patterns based approach, which describes significant changes between two classes of data, to recognize sequential, interleaved, and concurrent activities. Rashidi and Cook [39] proposed a stream mining method, which is extended from the tilted-time window approach [40], for automatically discovering human activity patterns over time from streaming sensor data.

Different from all the above supervised and unsupervised approaches, in this paper, we are not interested in what exactly each activity does (i.e., the label of each activity) since our goal is not to recognize elders' different activities. Instead, we attempt to identify elders' movements from a series of sensor reading because these reveal how they move in their smart homes, which enables us to reason the relationship between sensors so as to deduce a global topology. We develop PDS based on PMA [47] by further considering concurrent elders' movements, speculating elder routines, and validating PDS with two different use cases.

## 3. The methodology of PDS

The right part of Figure 1 shows the workflow of PDS to deduce elders' privacy in a smart home. First of all, PDS eavesdrops sensor data sent from the smart home and then preprocesses all sensor data that it has eavesdropped into a readable sensor log. Figure 2 shows the format of the sensor log. Each log record indicates the time point at which a sensor with its unique ID (which could be an IP address) is triggered. The above eavesdropping and preprocessing is possible and achievable if an attacker is physically close to a smart home and uses some sophisticated eavesdropping tools, such as [26], to collect sensor data from the smart home.

After that, PDS deduces the global sensor topology of the smart home by identifying all activities performed by the elders living in this smart home from the log, especially their movements. With the derived topology and the sensor log, PDS conducts a series of deductions based on Frequent Pattern Mining to reason sensor locations in the smart home. Finally, PDS speculates the elders' daily routines according to the derived sensor locations and the log. In the following, we will detail how PDS conducts sensor topology deduction, sensor location reasoning, and elder routine speculation.

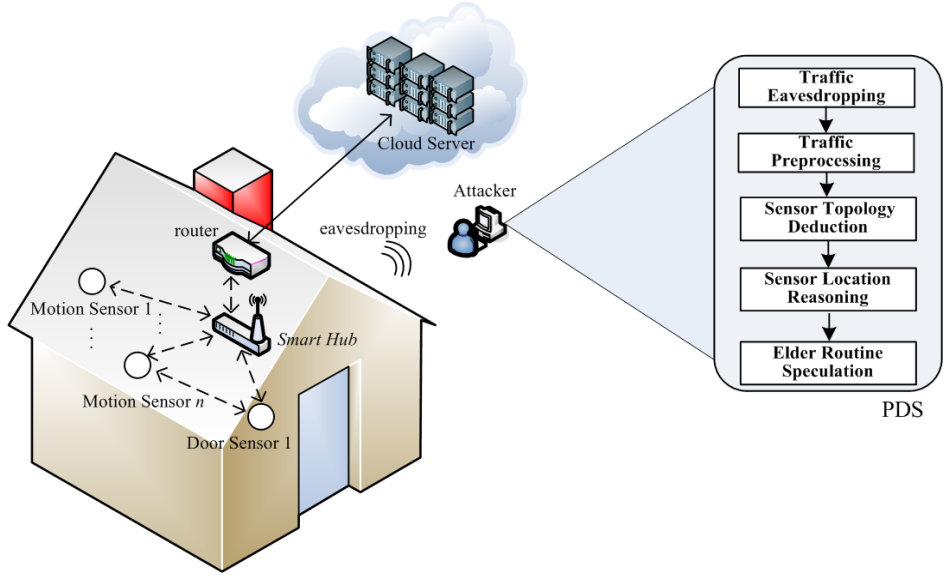

Figure 1. The workflow of PDS.

```
Date         Time       Sensor ID   Switch
2017-06-10   05:27:08   S013        ON
2017-06-10   05:27:12   S025        ON
2017-06-10   05:27:21   S019        ON
2017-06-10   05:27:24   S009        ON
2017-06-10   05:27:26   S017        ON
```

Figure 2. The format of sensor log.

### 3.1 Sensor Topology Deduction

This section describes how PDS deduces a global sensor topology for a smart home. To this purpose, we define two kinds of elder activities: *indoor activity* and *leave-back activity*. The former refers to an elder's movement in his/her house. Instead of identifying all kinds of activities, we are particularly interested in knowing how elders move in their homes since it enables us to infer how sensors are deployed in the homes. Therefore, in this paper, an indoor activity is an activity that lasts for a while and triggers a sequence of sensors. The rule to identify an indoor activity is defined as "*A sequence of sensors are triggered during a time period of at least $x$ seconds, and the interval between any two adjacent sensors is at most $y$ seconds where $x > y$*".

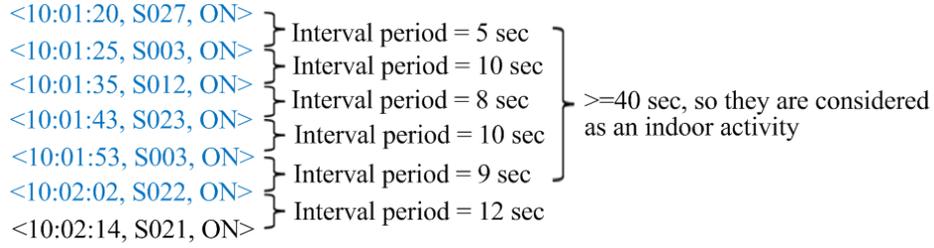

Figure 3. An example of indoor activity.

Taking Figure 3 as an example in which $x = 40$ sec and $y = 10$ sec, we know that the first six log records are considered as an indoor activity because their duration is 42 sec and the interval periods between any two adjacent records are no longer than 10 seconds. However, the last record is not considered as a part of the same activity since its interval period with the previous record is more than 10 seconds.

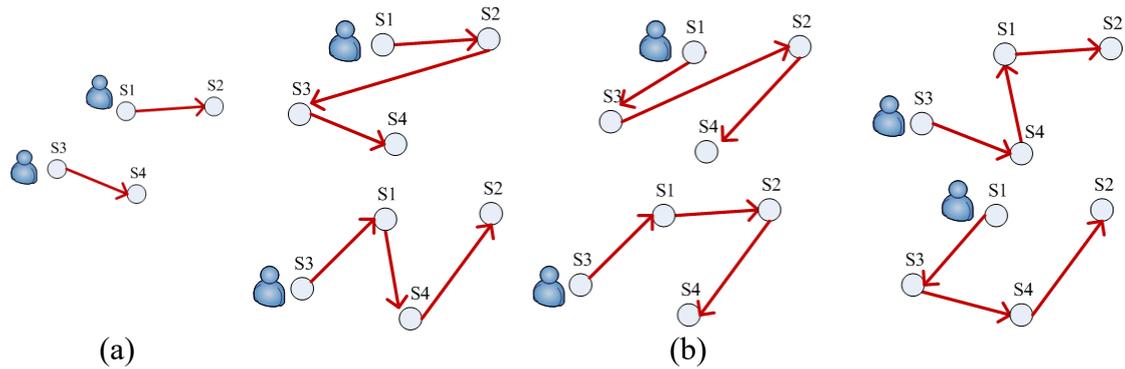

(a)                              (b)

Figure 4. An example showing how multiple elders' movements are falsely deduced as an indoor activity.

According to the above rule, it is clear that all identified indoor activities will not overlap with each other in terms of time. However, it is possible that multiple elders' concurrent movements are falsely identified as an indoor activity. For example, as illustrated in Figure 4(a), an elder moves from the kitchen to the living room (which triggers sensors S1 and S2) and meanwhile another elder moves from his bedroom to a toilet (which triggers sensors S3 and S4). The two movements might be falsely considered as an indoor activity if the definition rule holds, and its sensor-trigger sequence could be anyone of those depicted in Figure 4(b), which misleads PDS that a elder has such movement between these sensors and consequently misguides PDS in deducting a global sensor topology. In addition, it is also possible that two elders' movements that happen sequentially one after another are falsely identified as an indoor activity. These issues will be addressed and mitigated later in this paper.

On the other hand, a leave-back activity refers to the action that an elder leaves his/her house and then comes back after a while. When the elder leaves, a sensor deployed either on the door or near the house entrance will be triggered. The same sensor will be triggered again when the elder comes back. If there is no other in the house, no sensor in the house should be triggered during this time period. Hence, the rule to identify a leave-back activity is defined as "*A sensor is triggered at time point u and it is triggered again at time point v, where u < v. The duration between u and v must last at least z seconds, and no other sensors are triggered during this period.*" For example, the two log records shown in Figure 5 will be deduced as a leave-back activity if $z$ is 3600 seconds.

<08:38:06, S008, ON>  
<11:09:48, S008, ON> } > 3600 sec, so they are considered as a leave-back activity

Figure 5. An example of leave-back activity.

Based on the above rules, PDS identifies all indoor activities and leave-back activities from the sensor log. Note that when an elder goes to sleep or takes a nap for more than $z$ seconds, the sensor deployed in his/her bedroom will be triggered, and this action might be misjudged as a leave-back activity. This problem will be solved later in the paper.

After identifying all indoor activities and leave-back activities, PDS deduces a global sensor topology by translating each indoor activity into a set of directed edges. This set of directed edges not only shows the elders' movements between sensors to perform the activity, but also reveals the spatial relationship of these sensors in the smart home. Recall that the activity shown in Figure 3 is an indoor activity, which can be translated into the following set of directed edges. Clearly it shows that the elder moves from S027 to S003, S012, S023, S003, and S022 for performing the indoor activity.

S027→S003, S003→S012, S012→S023, S023→S003, S003→S022

After translating all identified indoor activities into corresponding directed-edge sets, PDS accumulates a confidence value $i$ for every directed edge $A→B$ (where $A$ and $B$ are two different sensors in the smart home, and $i$ is the total number of occurrences that $A{\rightarrow}B$ appears in all the directed-edge sets). The higher value of $i$, the more confidence that the elder can directly move from sensor $A$ to sensor $B$. Based on the following two rules, a global sensor topology can be constructed.

Rule 1: If both $A{\rightarrow}B$ and $B{\rightarrow}A$ have a confidence value larger than a predefined threshold $\alpha$, PDS is confident that the elders are *able* to directly move from $A$ to $B$ and vice versa. Hence, PDS adds a bidirectional solid edge between $A$ and $B$. Note that $\alpha = \left\lfloor \frac{\beta}{\gamma} \right\rfloor$ where $\beta$ is the summation of the confidence values of all the directed edges, and $\gamma$ is the total number of all the directed edges. In other words, $\alpha$ is the average confidence value of all the directed edges.

Rule 2: If only $A{\rightarrow}B$ or only $B{\rightarrow}A$ has a confidence value larger than $\alpha$, PDS presumes that the elder is *possible* to directly move from $A$ to $B$ and vice versa. In this case, PDS adds a bidirectional dash edge between $A$ and $B$.

Recall that multiple elders' concurrent movements and sequential movements might be separately identified as an indoor activity, which in turn falsely creates connections between sensors that in reality are not directly reachable for elders in a smart home. Employing Rule 1 helps to mitigate this problem. From the viewpoint of probability, it is unlikely that exactly the same two movements performed by multiple elders occur concurrently or sequentially very often, and it is also impossible that many falsely identified indoor activities have the same sensor-trigger sequence. Back to the previous example, it might be possible that an elder moves from the kitchen to the living room and another moves from his bedroom to the toilet at the same time. However, it is unlikely that such concurrent actions happen frequently and result in the same sensor-trigger sequence. Therefore, the side effect caused by falsely identified indoor activities could be significantly suppressed by Rule 1.

Apparently, Rule 2 is less strict than Rule 1, and it might not be helpful to suppress falsely identified indoor activities. However, it enables PDS to derive a global view of sensors since most sensors could be therefore connected together.

**The sensor-location deduction algorithm**
**Input**: All indoor activities of a smart home in $w$ days and global sensor topology
**Output**: Bedroom sensors, kitchen/dining room sensors, and entrances sensors.
**Procedure:**

```
1:   Let A be an empty set;
2:   for each indoor activity that occurred between 2 am and 6 am of the w days{
3:       Translate the activity into a sensor-ID list;
4:       Put the sensor-ID list into A;}
5:   Apply ARL with minSupport = 0.5 on A;
6:   Let i = 1;
7:   do {
8:       Choose a sensor set from the outcome of ARL if this set is the largest one and its
9:       occurrence is the most frequent compared with all other sets of the same size;
10:      Let this set be B and output B as sensors in bedroom i;
11:      while a sensor in the topology has bidirectional edges with at least a half of B {
12:          Consider this sensor to be in bedroom i;
13:          Put the sensor into B;}
14:      Discard any sensor-ID list that contains any sensor of B from A;
15:      Apply ARL with minSupport = 0.5 on A again;
16:      i = i + 1;
17:  }while ARL outputs at least one sensor set & the size of the set is larger than 1;
18:  Let C be an empty set;
19:  for each indoor activity that occurred between 6 pm and 7 pm of the w days{
20:      Translate the activity into a sensor-ID list;
21:      Put the sensor-ID list into C;}
22:  Apply ARL with minSupport = 0.5 on C;
23:  Choose a sensor set from the outcome of ARL if this set is the largest one and its
24:  occurrence is the most frequent compared with all other sets of the same size;
25:  Let this set be K and output K as sensors in kitchen/dining room;
26:  while a sensor in the topology has bidirectional edges with at least a half of K {
27:       Also consider this sensor as a sensor in the kitchen/dining room;
28:       Put the sensor into K;}
29:  Let e=1;
30:  for each leave-back activity{
31:     if the sensor in the activity does not equal to any deduced bedroom sensor{
32:        Consider it as an entrance sensor;
33:        if the sensor is the first deduced entrance sensor{
34:           Create an empty entrance-sensor set E_1 and put the sensor into E_1;
35:        }else{
36:          Let f = false;
37:          for j=1 to e {  //Note that e is the total number of entrance-sensor sets;
38:              if the sensor has bidirectional edges with at least a half of E_j {
39:                 Consider the sensor is close to the location of E_j;
40:                 Put the sensor into E_j;}
41:                 f = true; }}
42:          if f == false{
43:             Create a new entrance-sensor set and put the sensor into this set;
44:             e=e+1;}}}}
```

Figure 6. The sensor-location deduction algorithm.

### 3.2 Sensor Location Reasoning

Human are creatures of habit [41]. Everyone has his/her own daily routine and this routine mostly remain similar every day. For instance, an elder might wake up around 7 am everyday. If we monitor this elder long enough, we are able to see this phenomenon. Because of this, PDS needs to collect sensor data from a smart home and derives all the corresponding indoor activities for a sufficiently long period of time in order to be able to deduce sensor locations. Let $w$ be the total number of days in the period. In this paper, Association Rule Learning (ARL) [42] is used to find sensors associated with the same location. ARL is a rule-based machine learning method for finding groups of items that are commonly found together in a dataset, and it is typically used for market-basket analysis. In this method, the rule $X \Rightarrow Y$ holds with minSupport $s$ if at least $s * 100\%$ of a dataset contain both $X$ and $Y$ where $0 < s < 1$. In other words, both $X$ and $Y$ appear together in at least $s * 100\%$ of the dataset.

Figure 6 illustrates the algorithm utilized by PDS to deduce sensor locations. In this paper, we mainly focus on bedrooms, kitchen/dining room, and entrances. Deducing other places will be our future work. Note that here we do not differentiate kitchen and dining room because in many houses they are in the same place. PDS first deduces bedroom sensors (see lines 1 to 17) by reasonably assuming that most people stay in their bedrooms between 2 am and 6 am, implying that the sensors deployed in bedrooms are very likely to be triggered during this time period. PDS translates every indoor activity that occurred between 2 am and 6 am of these $w$ days into a sensor-ID list by extracting all distinct sensor IDs from the activity. For example, the sensor-ID list of the indoor activity shown in Figure 3 is {S027, S003, S012, S023, S022}. Let $A$ be a set of sensor-ID lists that associates with all indoor activities occurred between 2 am and 6 am of the $w$ days. PDS then applies ARL with minSupport of 0.5 on $A$ to find out all possible sets of sensors that satisfy the minSupport, i.e., the majority.

In order to deduce as much bedroom sensors as possible, PDS chooses a set when is the largest one and the occurrence of this set is the most frequent one as compared with all other sets of the same size (see lines 8 and 9). Because of the assumption, this set of sensors is therefore deduced as sensors in a same bedroom. Let $B$ be these sensors. After that, PDS attempts to deduce more sensors in the same bedroom by referring and inspecting the global sensor topology as follows: If any sensor in the topology has bidirectional edges with at least a half of $B$ based on majority rule, we believe that this is not a coincident caused by multiple elders' concurrent movements. Thus, PDS deduces that this sensor is also in the same bedroom. Furthermore, it is possible that a smart home has more than one bedroom. By removing all sensor-ID lists that contains any known bedroom sensor from $A$ and repeating the above steps, PDS is able to find sensors in other bedrooms.

Next, PDS deduces sensors in the kitchen/dining room (see lines 18 to 28) under the assumption that most people stay in their kitchens/dining rooms between 6 pm and 7 pm. Based on this assumption, PDS translates each indoor activity that occurred between this time period of the $w$ days into a sensor-ID list and repeats the same procedure (i.e., using ARL and checking the global sensor topology) to deduce all possible sensors in the kitchen/dining room. Due to the fact that ARL with the minSupport of 0.5 is employed, it is unlikely that a sensor that has been deduced as a bedroom sensor is also deduced as a kitchen/dining room sensor because, normally speaking, in most cases people do not stay in their bedrooms during 6 pm and 7 pm.

PDS continues by deducing entrance sensors (see lines 29 to 44). Recall that an indoor activity might be misjudged as a leave-back activity if an elder goes to sleep or takes a nap for longer than $z$ seconds. This issue could be addressed by elimination as follows: If a sensor has been deduced as a bedroom sensor, it is impossible to be an entrance sensor at the same time. Therefore, the sensor of a leave-back activity is confirmed to be an entrance sensor if it does not equal to any bedroom sensor that PDS has deduced. By further checking the global sensor topology (see lines 37 to 41), PDS can easily know how an entrance sensor is spatially related to other entrance sensors and deduce the total number of entrances in the smart home. In other words, PDS is able to deduce possible sensors at different entrances.

**3.3 Elder Routine Speculation**

With the knowledge of bedroom sensors, kitchen/dining room sensors, and entrance sensors, PDS is able to infer elders' daily routines in these places by using all the known sensors to mine the sensor log and meanwhile taking all identified indoor activities and leave-back activities into account. Our goal is to find as much as possible elders' activities in the above places, so we consider both crossing-sensor indoor activities and non-crossing-sensor indoor activities. The former type of activities refers to indoor activities that individually involve only one sensor. A typical example is that an elder cooks in the kitchen and only triggers one area sensor. The latter type of activities refers to indoor activities involving more than one sensor. A representative example is that an elder walks from one room to another room and therefore triggers several sensors. By accumulating all crossing-sensor indoor activities and non-crossing-sensor indoor activities occurred in bedrooms, PDS is able to know the frequency of elders' bedroom activities and the time periods these activities might happen. In other words, PDS knows when and how frequent the bedrooms of a smart home are used. Similarly, PDS knows the utilization pattern of the kitchen/dining room by accumulating all crossing-sensor indoor activities and non-crossing-sensor indoor activities occurred in kitchen/dining room. By analyzing all derived leave-back activities, PDS even knows when and how frequent the elders leave their houses.

## 4. Experimental Results

To evaluate the ability of PDS on deducing elders' privacy, we chose two real smart homes provided by WSU CASAS smart home project [43] to be our use cases. Table I shows the characteristics of the two smart homes. The first smart home "Milan" has only one floor, and the residents in this house are an old woman and a dog. The second smart home "Tulum" has two floors, and a couple live in this house. The reason we chose these two houses is that both of them have more than one resident, which might cause multiple indoor activities and therefore interfere PDS to deduce sensor relationships and sensor locations. Both smart homes have their own datasets recording the time at which a sensor is triggered. The dataset of Milan contains 433,665 sensor records generated by 7 area motion sensors, 21 motion sensors, 3 door closure sensors, and 2 temperature sensors. On the other hand, the dataset of Tulum contains 1,085,902 sensor records generated by 5 area motion sensors, 26 motion sensors, and 5 temperature sensors. In the following experiments, we ignored the sensor data generated by temperature sensors since this data is irrelevant. To simulate sensor traffic, we sequentially replayed each sensor record of these datasets and launched PDS.

Table I. The characteristics of two real smart homes.

| Name | Num. of Floors | Residents | Sensors |
|---|---|---|---|
| Milan | 1 | An old woman and a dog | 7 area motion sensors, 21 motion sensors, 3 door sensors, and 2 temperature sensors |
| Tulum | 2 | A couple | 5 area motion sensors, 26 motion sensors, and 5 temperature sensors |

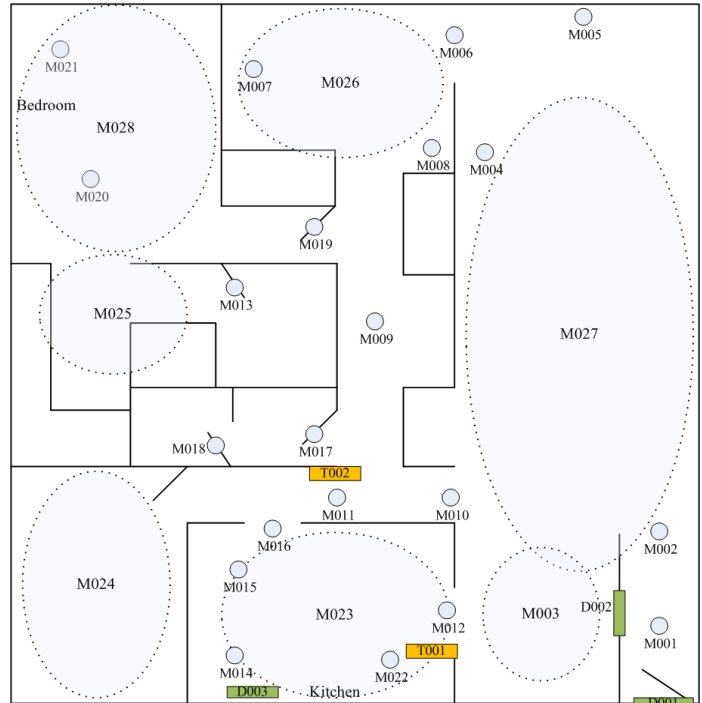

Figure 7. The Layout and sensor deployment of smart home Milan [44]. Note that the sensing coverage of an area motion sensor is represented by an oval. The sensing coverage of the other sensors is individually represented by a small circle.

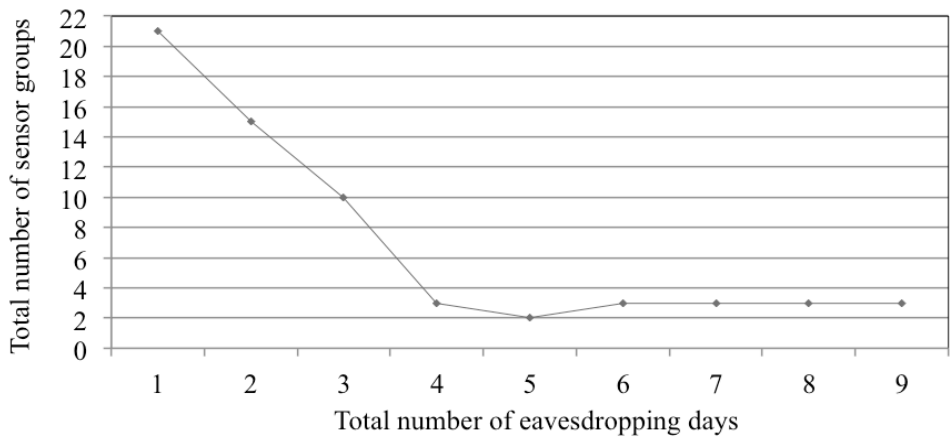

Figure 8. Total number of sensor groups in Milan deduced by PDS over time.

### 4.1 Case Study I: Milan

Figure 7 illustrates the layout and sensor deployment of Milan. Our goal is to see if PDS can correctly deduce a global sensor topology and infer sensors deployed in the bedrooms, kitchen/dining room, and entrances of Milan without knowing Figure 7. In this experiment, PDS eavesdropped Milan by using parameters $x = 40$ sec, $y = 10$ sec, and $z = 3600$ sec to identify all indoor activities and leave-back activities. The reason we adopted these parameters will be explained later in the discussion section. As shown in Figure 8, PDS was able to connect all the sensors that it found into 21 sensor groups after it eavesdropped Milan for just one day. Note that a sensor group is a group of sensors with directed edges connecting these sensors, implying that there is no directed edge between any two sensor groups. The number of sensor groups decreased to two on day 5, but it increased to three on day 6 because a new sensor appeared and was discovered by PDS. After day 6, the number of sensor groups remained identical as 3 and did not decrease any more. The reason is that the residents of Milan did not have lot of movements to certain places of the house, so the relevant sensors do not have sufficient directed edges with all other sensors.

By referring the number of sensor groups, we can determine an appropriate eavesdropping period for PDS to balance the trade-off between its deduction performance and the time effort required by an attacker to run PDS. As we can see from Figure 8, eavesdropping Milan for 6 days (sensor data was from Oct. 16$^{th}$ 2009 to Oct. 21$^{st}$ 2009) is sufficient for PDS and it is the most cost-efficient choice for attackers. Therefore, in the rest experiment, PDS utilized the sensor log of these six days to conduct its deduction.

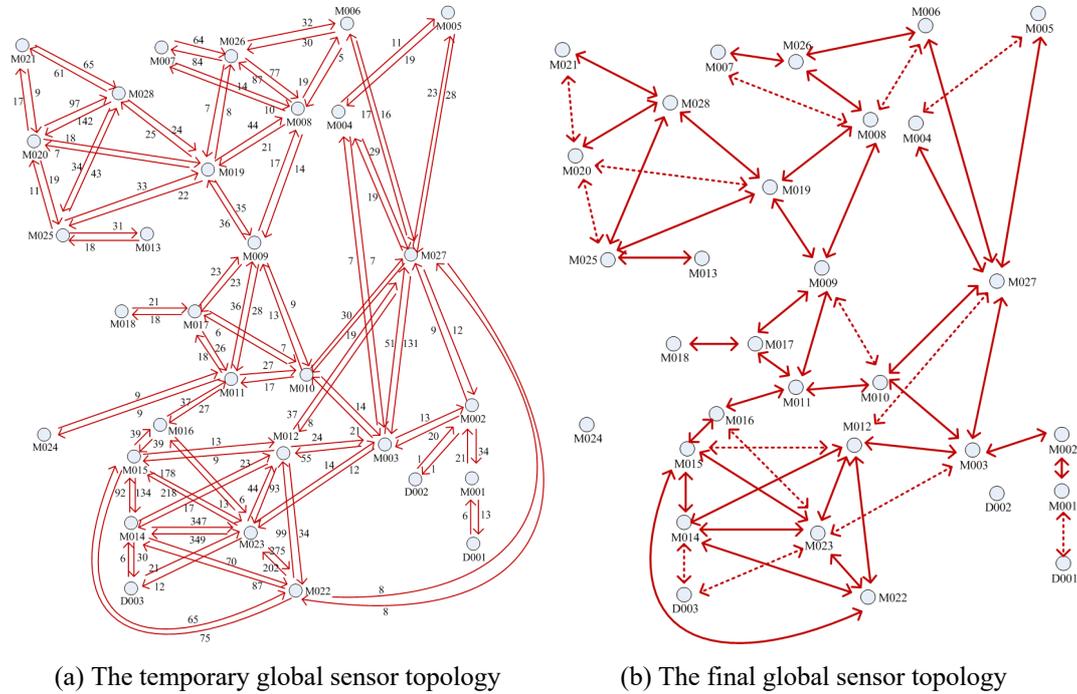

(a) The temporary global sensor topology      (b) The final global sensor topology

Figure 9. The global sensor topology of Milan.

Figure 9(a) shows the temporary global sensor topology in which a confidence value is presented next to each directed edge. A higher confidence value from sensor *A* to sensor *B* implies that PDS is more confident that an elder is able to directly move from sensor *A* to sensor *B*, where *A* and *B* are two different sensors in Milan. In the case of Milan, threshold $\alpha = 13 = \left\lceil \frac{5597}{415} \right\rceil$, so the final global sensor topology is the one illustrated in Figure 9(b) after Rule 1 and Rule 2 were applied. As compared with Figure 9(a), we can see that some edges disappear because their confidence values satisfy neither Rule 1 nor Rule 2. These eliminated edges are very likely caused by multiple residents' concurrent or sequential movements.

We can observe that PDS made some false deductions. For examples, there should be a bidirectional edge between M005 and M006 because the woman living in Milan should be able to directly move from M005 to M006 and vice versa according to the layout of Milan (see the right upper part of Figure 7). However, PDS was unable to deduce such relationship. Besides, from Figure 9(b), we can see that PDS was unable to deduce any relationship between M024 and the rest sensors. In other words, M024 itself is a sensor group. The key reason is that the confidence values between M024 and all other sensors are all lower than the predefined threshold (see Figure 9(a)), implying that the woman did not have a lot of movements to the place where M024 is deployed. The same situation happens for D002, so D002 itself is also a sensor group.

| From\To | M001 | M002 | M003 | M004 | M005 | M006 | M007 | M008 | M009 | M010 | M011 | M012 | M013 | M014 | M015 | M016 | M017 | M018 | M019 | M020 | M021 | M022 | M023 | M024 | M025 | M026 | M027 | M028 | D001 | D002 | D003 |
|---|---|---|---|---|---|---|---|---|---|---|---|---|---|---|---|---|---|---|---|---|---|---|---|---|---|---|---|---|---|---|---|
| M001 | -- | 1/1 | 0/0 | 0/0 | 0/0 | 0/0 | 0/0 | 0/0 | 0/0 | 0/0 | 0/0 | 0/0 | 0/0 | 0/0 | 0/0 | 0/0 | 0/0 | 0/0 | 0/0 | 0/0 | 0/0 | 0/0 | 0/0 | 0/0 | 0/0 | 0/0 | 0/0 | 0/0 | 1/1 | **1/0** | 0/0 |
| M002 | 1/1 | -- | 1/1 | 0/0 | 0/0 | 0/0 | 0/0 | 0/0 | 0/0 | 0/0 | 0/0 | 0/0 | 0/0 | 0/0 | 0/0 | 0/0 | 0/0 | 0/0 | 0/0 | 0/0 | 0/0 | 0/0 | 0/0 | 0/0 | 0/0 | **1/0** | 0/0 | 0/0 | **1/0** | 0/0 |
| M003 | 0/0 | 1/1 | -- | 0/0 | 0/0 | 0/0 | 0/0 | 0/0 | 0/0 | 1/1 | 0/0 | 1/1 | 0/0 | 0/0 | 0/0 | 0/0 | 0/0 | 0/0 | 0/0 | 0/0 | 0/0 | 0/0 | 0/0 | 0/0 | 0/0 | 1/1 | 0/0 | 0/0 | 0/0 | 0/0 |
| M004 | 0/0 | 0/0 | 0/0 | -- | 1/0 | 0/0 | 0/0 | 0/0 | 0/0 | 0/0 | 0/0 | 0/0 | 0/0 | 0/0 | 0/0 | 0/0 | 0/0 | 0/0 | 0/0 | 0/0 | 0/0 | 0/0 | 0/0 | 0/0 | 0/0 | 1/1 | 0/0 | 0/0 | 0/0 | 0/0 |
| M005 | 0/0 | 0/0 | 0/0 | 1/1 | -- | 1/0 | 0/0 | 0/0 | 0/0 | 0/0 | 0/0 | 0/0 | 0/0 | 0/0 | 0/0 | 0/0 | 0/0 | 0/0 | 0/0 | 0/0 | 0/0 | 0/0 | 0/0 | 0/0 | 0/0 | 0/0 | 0/0 | 0/0 | 0/0 | 0/0 |
| M006 | 0/0 | 0/0 | 0/0 | 0/0 | 1/0 | -- | 0/0 | 1/0 | 0/0 | 0/0 | 0/0 | 0/0 | 0/0 | 0/0 | 0/0 | 0/0 | 0/0 | 0/0 | 0/0 | 0/0 | 0/0 | 0/0 | 0/0 | 0/0 | 1/1 | 1/1 | 0/0 | 0/0 | 0/0 | 0/0 |
| M007 | 0/0 | 0/0 | 0/0 | 0/0 | 0/0 | 0/0 | -- | 1/1 | 0/0 | 0/0 | 0/0 | 0/0 | 0/0 | 0/0 | 0/0 | 0/0 | 0/0 | 0/0 | 0/0 | 0/0 | 0/0 | 0/0 | 0/0 | 0/0 | 0/0 | 1/1 | 0/0 | 0/0 | 0/0 | 0/0 |
| M008 | 0/0 | 0/0 | 0/0 | 0/0 | 1/1 | 1/0 | -- | 1/1 | 0/0 | 0/0 | 0/0 | 0/0 | 0/0 | 0/0 | 0/0 | 1/1 | 0/0 | 0/0 | 0/0 | 0/0 | 0/0 | 0/0 | 0/0 | 1/1 | 0/0 | 0/0 | 0/0 | 0/0 | 0/0 | 0/0 |
| M009 | 0/0 | 0/0 | 0/0 | 0/0 | 0/0 | 0/0 | 0/0 | 1/1 | -- | 1/0 | 1/1 | 0/0 | 0/0 | 0/0 | 0/0 | 1/0 | 1/1 | 0/0 | 0/0 | 0/0 | 0/0 | 0/0 | 0/0 | 0/0 | 0/0 | 0/0 | 0/0 | 0/0 | 0/0 | 0/0 |
| M010 | 0/0 | 0/0 | 1/1 | 0/0 | 0/0 | 0/0 | 0/0 | 0/0 | 1/1 | -- | 1/1 | 0/0 | 0/0 | 0/0 | 0/0 | 0/0 | 0/0 | 0/0 | 0/0 | 0/0 | 0/0 | 0/0 | 0/0 | 0/0 | 1/1 | 0/0 | 0/0 | 0/0 | 0/0 | 0/0 |
| M011 | 0/0 | 0/0 | 0/0 | 0/0 | 0/0 | 0/0 | 0/0 | 0/0 | 1/1 | 1/1 | -- | 0/0 | 0/0 | 0/0 | 0/0 | 1/1 | 1/1 | 0/0 | 0/0 | 0/0 | 0/0 | 0/0 | 0/0 | 0/0 | **1/0** | 0/0 | 0/0 | 0/0 | 0/0 | 0/0 |
| M012 | 0/0 | 0/0 | 1/1 | 0/0 | 0/0 | 0/0 | 0/0 | 0/0 | 0/0 | 0/0 | 0/0 | -- | 0/0 | 1/1 | **1/0** | 0/0 | 0/0 | 0/0 | 0/0 | 0/0 | 0/0 | 1/1 | 1/1 | 0/0 | 0/0 | 1/1 | 0/0 | 0/0 | 0/0 | 0/0 |
| M013 | 0/0 | 0/0 | 0/0 | 0/0 | 0/0 | 0/0 | 0/0 | 0/0 | 0/0 | 0/0 | 0/0 | 0/0 | -- | 0/0 | 0/0 | 0/0 | 0/0 | 0/0 | 0/0 | 0/0 | 0/0 | 0/0 | 0/0 | 1/1 | 0/0 | 0/0 | 0/0 | 0/0 | 0/0 | 0/0 |
| M014 | 0/0 | 0/0 | 0/0 | 0/0 | 0/0 | 0/0 | 0/0 | 0/0 | 0/0 | 0/0 | 0/0 | 1/1 | 0/0 | -- | 1/1 | 0/0 | 0/0 | 0/0 | 0/0 | 0/0 | 0/0 | 1/1 | 1/1 | 0/0 | 0/0 | 0/0 | 0/0 | 0/0 | 0/0 | 0/0 | 1/1 |
| M015 | 0/0 | 0/0 | 0/0 | 0/0 | 0/0 | 0/0 | 0/0 | 0/0 | 0/0 | 0/0 | 0/0 | 1/1 | 0/0 | 1/1 | -- | 1/1 | 0/0 | 0/0 | 0/0 | 0/0 | 0/0 | 1/1 | 1/1 | 0/0 | 0/0 | 0/0 | 0/0 | 0/0 | 0/0 | 0/0 | 0/0 |
| M016 | 0/0 | 0/0 | 0/0 | 0/0 | 0/0 | 0/0 | 0/0 | 0/0 | 1/1 | 0/0 | 0/0 | 0/0 | 1/1 | -- | 0/0 | 0/0 | 0/0 | 0/0 | 0/0 | 0/0 | 0/0 | **0/0** | **1/0** | 0/0 | 0/0 | 0/0 | 0/0 | 0/0 | 0/0 | 0/0 |
| M017 | 0/0 | 0/0 | 0/0 | 0/0 | 0/0 | 0/0 | 0/0 | 1/1 | 0/0 | 1/1 | 0/0 | 0/0 | 0/0 | 0/0 | 0/0 | -- | 1/1 | 0/0 | 0/0 | 0/0 | 0/0 | 0/0 | 0/0 | 0/0 | 0/0 | 0/0 | 0/0 | 0/0 | 0/0 | 0/0 |
| M018 | 0/0 | 0/0 | 0/0 | 0/0 | 0/0 | 0/0 | 0/0 | 0/0 | 0/0 | 0/0 | 0/0 | 0/0 | 0/0 | 0/0 | 0/0 | 0/0 | 1/1 | -- | 0/0 | 0/0 | 0/0 | 0/0 | 0/0 | 0/0 | 0/0 | 0/0 | 0/0 | 0/0 | 0/0 | 0/0 | 0/0 |
| M019 | 0/0 | 0/0 | 0/0 | 0/0 | 0/0 | 0/0 | 0/0 | 1/1 | 1/1 | 0/0 | 0/0 | 0/0 | 0/0 | 0/0 | 0/0 | 0/0 | 0/0 | 0/0 | -- | 0/0 | 0/0 | 0/0 | 0/0 | 1/1 | 0/0 | 0/0 | 1/1 | 0/0 | 0/0 | 0/0 | 0/0 |
| M020 | 0/0 | 0/0 | 0/0 | 0/0 | 0/0 | 0/0 | 0/0 | 0/0 | 0/0 | 0/0 | 0/0 | 0/0 | 0/0 | 0/0 | 0/0 | 0/0 | 0/0 | 0/0 | 1/1 | -- | 1/1 | 0/0 | 0/0 | 1/1 | 0/0 | 0/0 | 1/1 | 0/0 | 0/0 | 0/0 | 0/0 |
| M021 | 0/0 | 0/0 | 0/0 | 0/0 | 0/0 | 0/0 | 0/0 | 0/0 | 0/0 | 0/0 | 0/0 | 0/0 | 0/0 | 0/0 | 0/0 | 0/0 | 0/0 | 0/0 | 0/0 | **1/0** | -- | 0/0 | 0/0 | 0/0 | 0/0 | 0/0 | 1/1 | 0/0 | 0/0 | 0/0 | 0/0 |
| M022 | 0/0 | 0/0 | 0/0 | 0/0 | 0/0 | 0/0 | 0/0 | 0/0 | 0/0 | 0/0 | 0/0 | 1/1 | 0/0 | 1/1 | 1/1 | 0/0 | 0/0 | 0/0 | 0/0 | 0/0 | 0/0 | -- | 0/0 | 0/0 | 0/0 | 0/0 | 0/0 | 0/0 | 0/0 | 0/0 | 0/0 |
| M023 | 0/0 | 0/0 | **0/1** | 0/0 | 0/0 | 0/0 | 0/0 | 0/0 | 0/0 | 0/0 | 0/0 | 1/1 | 0/0 | 1/1 | 1/1 | 1/1 | 0/0 | 0/0 | 0/0 | 0/0 | 0/0 | 1/1 | -- | 0/0 | 0/0 | 1/1 | 0/0 | 0/0 | 0/0 | 0/0 | **1/0** |
| M024 | 0/0 | 0/0 | 0/0 | 0/0 | 0/0 | 0/0 | 0/0 | 0/0 | 0/0 | **1/0** | 0/0 | 0/0 | 0/0 | 0/0 | 0/0 | 0/0 | 0/0 | 0/0 | 0/0 | 0/0 | 0/0 | 0/0 | 0/0 | -- | 0/0 | 0/0 | **1/0** | 0/0 | 0/0 | 0/0 | 0/0 |
| M025 | 0/0 | 0/0 | 0/0 | 0/0 | 0/0 | 0/0 | 0/0 | 0/0 | 0/0 | 0/0 | 1/0 | 0/0 | 0/0 | 0/0 | 0/0 | 1/1 | 0/0 | 0/0 | 1/1 | 11 | 0/0 | 0/0 | 0/0 | 0/0 | -- | 0/0 | 0/0 | 1/1 | 0/0 | 0/0 | 0/0 |
| M026 | 0/0 | 0/0 | 0/0 | 0/0 | 0/0 | 0/0 | 1/1 | 1/1 | 1/1 | 0/0 | 0/0 | 0/0 | 0/0 | 0/0 | 0/0 | 0/0 | 0/0 | 0/0 | 0/0 | 0/0 | 0/0 | 0/0 | 0/0 | 0/0 | 0/0 | -- | 0/0 | 0/0 | 0/0 | 0/0 | 0/0 |
| M027 | 0/0 | **1/0** | 1/1 | 1/1 | 1/1 | 1/1 | 0/0 | 0/0 | 0/0 | 1/1 | 0/0 | **1/0** | 0/0 | 0/0 | 0/0 | 0/0 | 0/0 | 0/0 | 0/0 | 0/0 | 0/0 | 0/0 | **1/0** | 0/0 | 0/0 | 0/0 | -- | 0/0 | 0/0 | 0/0 | 0/0 |
| M028 | 0/0 | 0/0 | 0/0 | 0/0 | 0/0 | 0/0 | 0/0 | 0/0 | 0/0 | 0/0 | 0/0 | 0/0 | 0/0 | 0/0 | 0/0 | 0/0 | 0/0 | 1/1 | 1/1 | 1/1 | 0/0 | 0/0 | 0/0 | 0/0 | 1/1 | 0/0 | 0/0 | -- | 0/0 | 0/0 | 0/0 |
| D001 | 1/0 | 0/0 | 0/0 | 0/0 | 0/0 | 0/0 | 0/0 | 0/0 | 0/0 | 0/0 | 0/0 | 0/0 | 0/0 | 0/0 | 0/0 | 0/0 | 0/0 | 0/0 | 0/0 | 0/0 | 0/0 | 0/0 | 0/0 | 0/0 | 0/0 | 0/0 | 0/0 | 0/0 | -- | 0/0 | 0/0 |
| D002 | 1/0 | **1/0** | 0/0 | 0/0 | 0/0 | 0/0 | 0/0 | 0/0 | 0/0 | 0/0 | 0/0 | 0/0 | 0/0 | 0/0 | 0/0 | 0/0 | 0/0 | 0/0 | 0/0 | 0/0 | 0/0 | 0/0 | 0/0 | 0/0 | 0/0 | 0/0 | 0/0 | 0/0 | 0/0 | -- | 0/0 |
| D003 | 0/0 | 0/0 | 0/0 | 0/0 | 0/0 | 0/0 | 0/0 | 0/0 | 0/0 | 0/0 | 0/0 | 0/0 | 1/1 | 0/0 | 0/0 | 0/0 | 0/0 | 0/0 | 0/0 | 0/0 | 0/0 | 0/0 | 1/1 | 0/0 | 0/0 | 0/0 | 0/0 | 0/0 | 0/0 | 0/0 | -- |

Figure 10. The sensor spatial relationships of Milan and those deduced by PDS. Note that all false deductions are highlighted.

Figure 10 shows the inference accuracy of PDS on the sensor relationships of Milan where

1. 1/1 in attribute (*A*/*B*) means that sensor *B* is directly reachable from sensor *A* according to the layout of Milan, and PDS is able to deduce a direct edge from *A* to *B* (i.e., $A \rightarrow B$).

2. 1/0 in attribute (*A*/*B*) means that *B* is directly reachable from *A* based on the layout, but PDS is unable to deduce direct edge $A \rightarrow B$.

3. 0/0 means that it is unable to directly reach from $A$ to $B$ based on the layout, and PDS is also unable to deduce direct edge $A \rightarrow B$.

4. 0/1 means that it is unable to directly reach from $A$ to $B$ based on the layout, but PDS is able to deduce direct edge $A \rightarrow B$.

It is clear that PDS makes an incorrect deduction when 1/0 or 0/1 appears. Recall that there are 31 sensors in Milan, so the total number of sensor-relationship deductions is 930 (=31x31-31). The total number of false deductions is 23 (see all the highlights in Figure 10), implying that the accuracy rate of PDS on sensor-relationship inference is 97.5% $\cong \left(1 - \frac{23}{930}\right) * 100\%$.

Next, PDS used the algorithm shown in Figure 6 to deduce sensors in the bedrooms, kitchen/dining room, and entrances of Milan. PDS first extracted all indoor activities between 2 am and 6 am from each of the 6 days. The total number of such activities is four, and the corresponding sensor-ID lists are shown in Figure 11. After applying ARL with minSupport of 0.5 on the four sensor-ID lists, PDS returned a sensor set {M013, M020, M021, M025, M028} because 1) the occurrence ratio of this set in the four lists (i.e., 2/4) satisfies the minSupport, 2) this set is the largest one, and 3) this set has the highest number of occurrences as compared with other sets with the same size. Therefore, this set of sensors was deduced as sensors in a bedroom. Since all other sensors in the topology except M019 do not have bidirectional edges with a half of this set, M019 was deduced as a sensor in the same bedroom. After that, PDS attempted to deduce sensors in another bedroom by discarding any sensor-ID list that contains any deduce bedroom sensors (i.e., M013, M019, M020, M021, M025, and M028). The resulting sensor-ID list is empty, implying that there is no other bedroom in Milan, so PDS stopped such deduction. As compared with the sensor deployment shown in Figure 7, we can see that PDS made a correct deduction, i.e., there is only one bedroom in Milan, and M013, M019, M020, M021, M025, and M028 are indeed deployed in the bedroom.

| 1 | M013 M020 M021 M025 M028 |
| 2 | M013 M020 M021 M025 M028 |
| 3 | M020 M021 M025 M028 |
| 4 | M020 M025 M028 |

Figure 11. All sensor-ID lists between 2 am and 6 am of the entire eavesdropping period for deducing bedroom sensors in Milan.

To deduce kitchen/dining room sensors, PDS extracted all indoor activities between 6 pm and 7 pm from each of the 6 days. Figure 12 lists all the corresponding sensor-ID lists. After applying ARL with minSupport of 0.5 on these lists, PDS returned a matching sensor set {M014, M015, M022, M023}. Hence, they were deduced as sensors deployed in the kitchen/dining room of Milan. Furthermore, since only D003, M012, and M016 in the global topology have bidirectional edges with a half of {M014, M015, M022, M023}, these three sensors were all deduced as sensors in the kitchen/dining room as well. By comparing the above result with the layout of Milan depicted in Figure 7, we know that this deduction is accurate.

```
1    M012 M014 M015 M022 M023
2    M014 M015 M022 M023
3    M008 M009 M011 M014 M015 M016 M017 M022 M023 M026
4    M005 M006 M008 M009 M011 M013 M019 M025 M026
5    M011 M014 M015 M016 M017 M018 M022 M023
6    M003 M005 M007 M008 M009 M011 M012 M014 M015 M016 M023 M026
7    M009 M011 M014 M015 M016 M022 M023
8    M007 M008 M009 M011 M014 M015 M016 M019 M022 M023 M026
9    M008 M009 M014 M015 M016 M017 M022
10   M014 M015 M022 M023
11   M014 M015 M022 M023
12   M014 M015 M023
13   M003 M004 M012 M015 M022 M023 M027
14   D003 M012 M014 M015 M022 M023
15   M005 M006 M007 M008 M026 M027
16   M012 M014 M022 M023
17   M014 M022 M023
18   M001 M002 M003 M012 M014 M023
19   D003 M012 M014 M023
20   M001 M002 M003 M012 M014 M015 M022 M023 M027
21   M002 M003 M004 M027
22   M012 M014 M015 M022 M023
23   D003 M012 M014 M015 M022 M023
24   M014 M015 M023
25   M014 M015 M023
26   M014 M015 M022 M023
27   M012 M014 M015 M022 M023
28   M014 M015 M022 M023
29   D003 M011 M014 M015 M016 M017 M022 M023
30   M001 M014 M015 M022 M023
31   M001 M002 M003 M012 M014 M022 M023 M027
32   M001 M002 M003 M006 M007 M008 M009 M010 M012 M014 M022 M023 M026 M027
33   M012 M014 M015 M022 M023
34   M003 M012 M014 M015 M022 M023
35   M014 M022 M023
36   M012 M014 M015 M022 M023
37   M014 M015 M022 M023
38   D003 M014 M015 M022 M023
39   M014 M022 M023
40   M001 M002 M003 M014 M015 M022 M023
41   M012 M022 M023
```

Figure 12. All sensor-ID lists between 6 pm and 7 pm of the entire eavesdropping period for deducing kitchen/dining room sensors in Milan.

PDS continued by deducing entrance sensors in Milan. To do so, PDS extracted all leave-back activities from the same time period. There are only two sensors in these activities: M028 and D001. Recall that M028 was already deduced as a bedroom sensor, so it could not be an entrance sensor at the same time. In other words, only D001 is a sensor at the entrance of Milan and this deduction is correct according to Figure 7. Note that from the bottom part of Figure 7, we can see that there is a door sensor called D003, but PDS was unable to find it since the woman living in Milan did not leave the house for more than one hour from the door where D003 is deployed according to the sensor log.

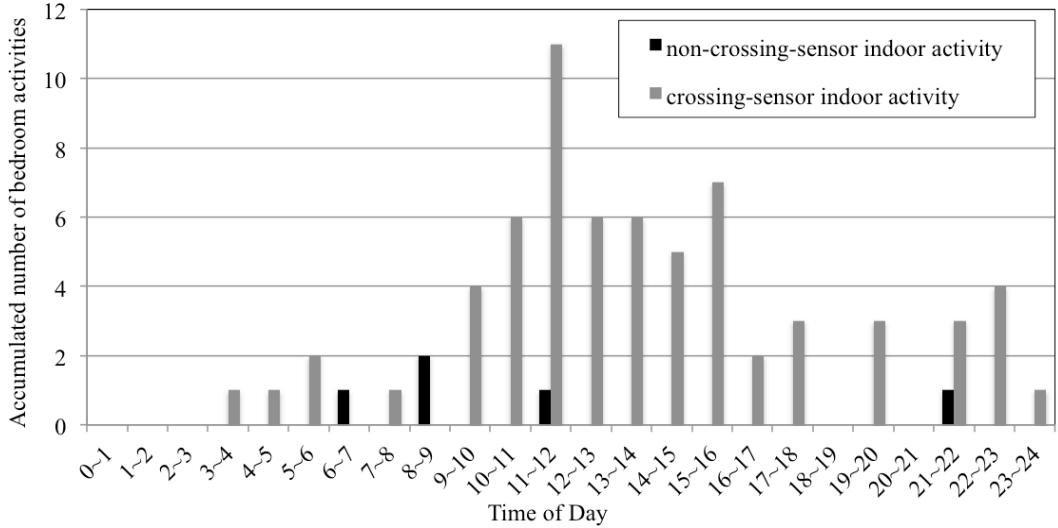

Figure 13. The accumulated number of indoor activities in Milan's bedroom during the entire eavesdropping period.

Figure 13 shows the accumulated number of crossing-sensor indoor activities and non-crossing-sensor indoor activities occurred in Milan's bedroom during the eavesdropping period. As an attacker, we can observe the following patterns: (1) No activity was found during 00:00 to 03:00 of the entire eavesdropping period, implying that the woman living in Milan may fall into a deep sleep during this period; (2) Starting from 03:00, there were some movements, including crossing-sensor and non-crossing-sensor activities, and the number of movements kept increasing after 07:00, implying that the woman might get up after 7 am; (3) Many crossing-sensor indoor activities were found in the bedroom during the daytime, especially between 11:00 and 12:00. Therefore, we know that the woman frequently stays in the bedroom during this period; (4) The woman is very likely to sleep between 21:00 and 00:00 since there was no activity in the bedroom before and after this period.

Figure 14 illustrates the accumulated number of crossing-sensor indoor activities and non-crossing-sensor indoor activities occurred in Milan's kitchen/dining room during the eavesdropping period. The results show that it is very likely that the woman's breakfast time is after 08:00 because no activity was observed before 08:00. Furthermore, her dinner time might be between 17:00 and 19:00 since lots of movements were found in the kitchen/dining room during this period. This statement can be further supported because the residents' movements in the kitchen/dining room dramatically reduced after 19:00.

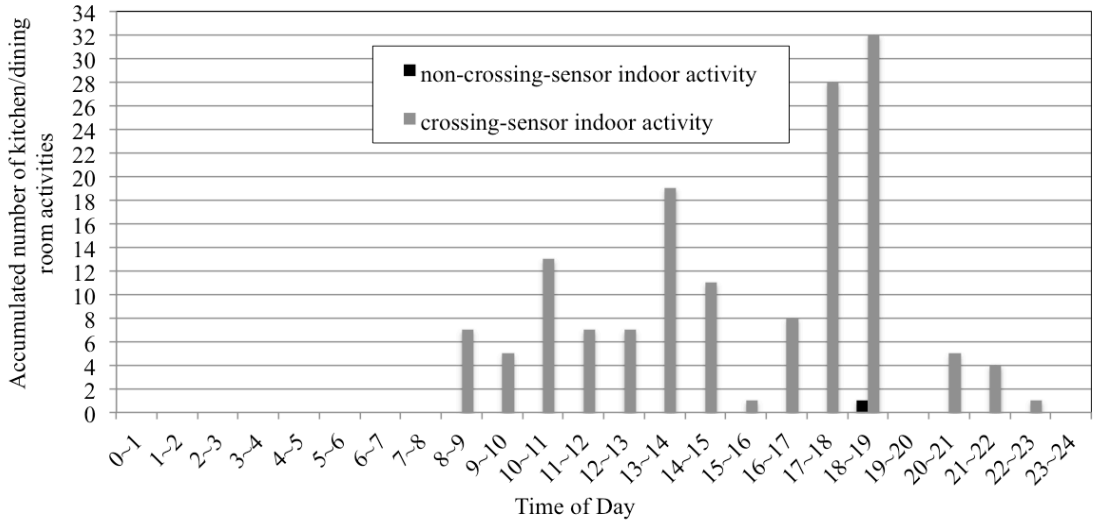

Figure 14. The accumulated number of indoor activities in Milan's kitchen/dining room during the entire eavesdropping period.

During the entire eavesdropping period, PDS found only one leave-back activity occurred from 12:32 to 14:43 on Oct. 16[th] when $z = 3600$ sec. Therefore, we can deduce that the woman usually stays at home, and she did not frequently go out together with her dog for more than one hour.

Based on the above deduction results, we confirm that PDS is able to expose a lot of information about Milan, including the global sensor topology, sensors in several places, and most importantly the woman's daily pattern.

**4.2 Cast Study II: Tulum**

In the section, we show the deduction ability of PDS on Tulum without knowing the layout and sensor deployments of Tulum depicted in Figure 15. PDS employed the same parameters that it used for Milan to identify all indoor activities and leave-back activities that occurred in Tulum by replaying and eavesdropping the sensor dataset of Tulum. The first date of the data is Sept. 27[th] 2009. As we can see from Figure 16, the number of sensor groups decreases to one after PDS eavesdropped Tulum for six days. Note that the sensor data was from Sept. 27[th] 2009 to Oct. 3[rd] 2009 because there was no sensor recorded on Oct. 2[nd] in the dataset of Tulum. To minimize an attacker's effort, these six days are sufficient for him/her to understand Tulum and its residents.

Therefore, in the following experiment, PDS used the sensor log of these six days to conduct its deduction.

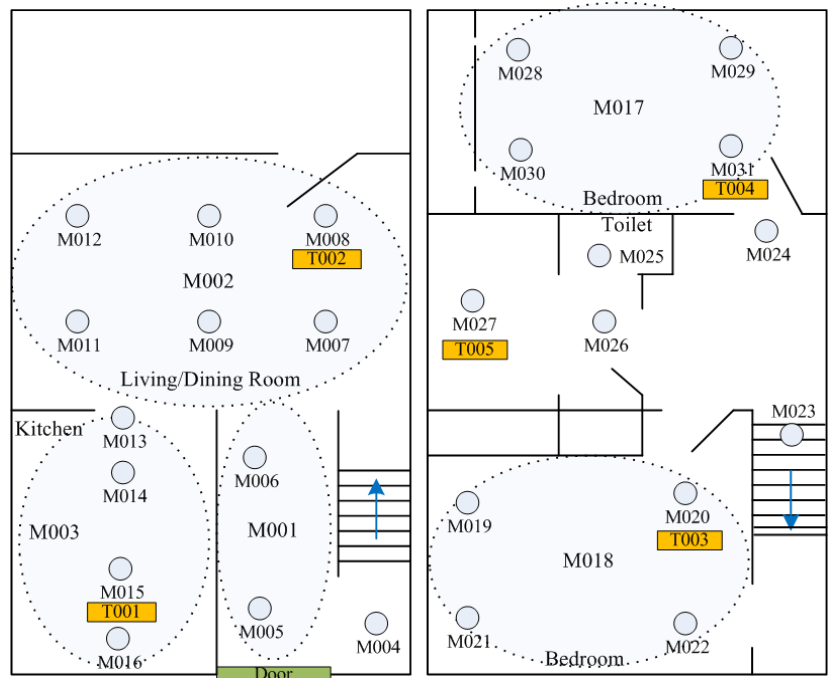

Figure 15. The layout and sensor deployment of smart home Tulum [22]. Note that the sensing coverage of an area motion sensor is represented by an oval. The sensing coverage of the other sensors is individually represented by a circle.

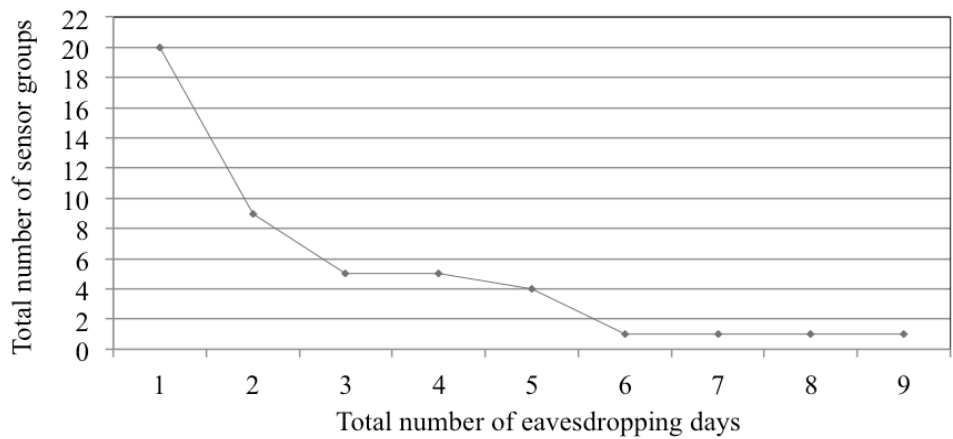

Figure 16. Total number of sensor groups of Tulum deduced by PDS over time.

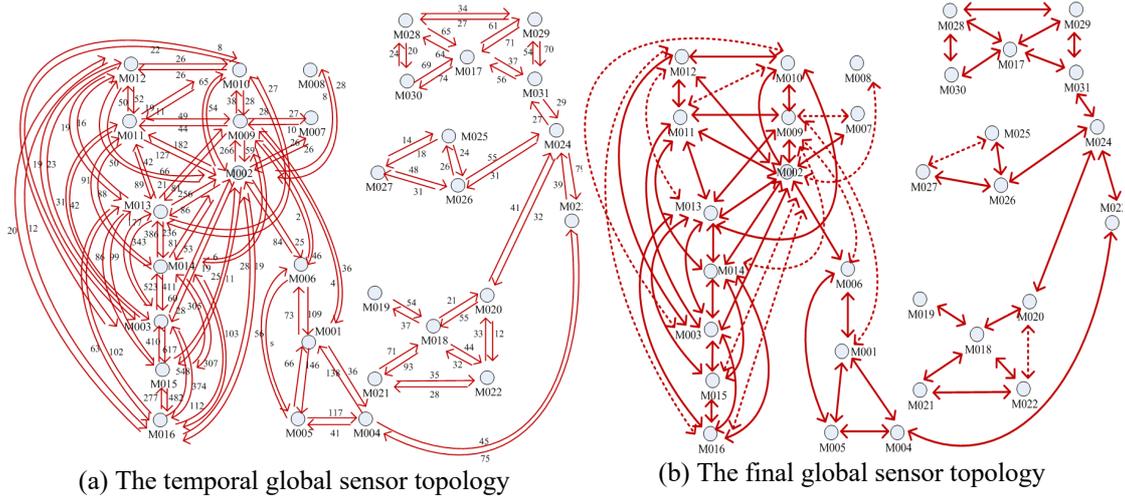

(a) The temporal global sensor topology  (b) The final global sensor topology

Figure 17. The global sensor topology of Tulum.

Figure 18. The sensor spatial relationships of Tulum and those deduced by PDS.

Figures 17(a) illustrates the temporal global sensor topology with all directed edges between sensors and the corresponding confidence values. In the case of Tulum, $\alpha = 18 = \left\lfloor \frac{13645}{728} \right\rfloor$ and hence the resulting final sensor topology is the one shown in Figure 17(b). Note that we can see the advantage of Rule 2 since this rule enables showing that the couple living in Tulum are able to directly move from M025 to M027 and vice versa. Such statement is correct according to Figure 15. Figure 18 shows the sensor spatial relationships of Tulum and those deduced by PDS. The total number of false deductions is 31, meaning that the inference accuracy of PDS on sensor relationships is $96.7\% \cong \left(1 - \frac{31}{930}\right) * 100\%$.

Next, PDS used the sensor-location deduction algorithm to deduce sensor locations. To deduce bedroom sensors in Tulum, PDS extracted all indoor activities that occurred between 2 am and 6 am from each of the six days. The corresponding sensor-ID lists are shown in Figure 19. After applying ARL with minSupport of 0.5 on these lists, PDS returned set {M018, M020, M022, M026}, which was therefore deduced as sensors in a bedroom. Since all other sensors in the topology except M021 have no bidirectional edges with a half of set {M018, M020, M022, M026}, M021 was also considered as a sensor deployed in the same bedroom. As compared with Figure 15, we can see that M018, M020, M021, and M022 are indeed in the same bedroom, but not M026. Nevertheless, a careful attacker might be able to notice from the global sensor topology that no directed edge exists between M026 and those bedroom sensors, so he/she knows that M026 is not deployed in the bedroom. In addition, the above deduction shows that PDS was unable to find that M019 is also in the same bedroom since the couple living in Tulum did not have lots of movements involving M019 during the eavesdropping period.

PDS continued inferring sensors in other bedrooms by discarding any sensor-ID list that contains M018, M020, M021, M022, or M026. The remaining sensor-ID list is only one as listed in Figure 20. Apparently, after applying ARL on this list, the returned sensor set will be the same, i.e., {M017, M029, M030, M031}, which was therefore deduced as sensors deployed in another bedroom. Moreover, as we can observe from the global sensor topology that only M028 has bidirectional edges with a half of this set, so M028 was also deduced as a sensor deployed in the same bedroom. In other words, sensors M017, M028, M029, M030, and M031 are all in the second bedroom. By verifying the above deduction results with Figure 15 again, we confirm that the deduction is correct. Due to the fact that the list shown in Figure 20 contains four out of the five sensors, PDS continues deducing sensors in another bedroom by discarding all founded bedroom sensors from the list shown in Fig. 20. Since the list becomes empty, PDS stopped deducing bedroom sensors. In summary, PDS found two bedrooms in Tulum and most sensors that are deployed in these two bedrooms.

| 1 | M018 M020 M021 M022 M025 M026 |
|---|---|
| 2 | M001 M002 M003 M004 M005 M006 M011 M013 M020 M021 M022 M023 M024 |
| 3 | M017 M029 M030 M031 |
| 4 | M018 M019 M020 M021 M022 M025 M026 M027 |
| 5 | M018 M019 M021 M030 |
| 6 | M018 M020 M021 M022 M024 M025 M026 |
| 7 | M018 M019 M020 M022 M024 M026 |

Figure 19. All sensor-ID lists between 2 am and 6 am of the entire eavesdropping period for deducing bedroom sensors in Tulum.

| 3 | M017 M029 M030 M031 |
|---|---|

Figure 20. The remaining sensor-ID lists for deducing sensors in another bedroom of Tulum.

```
1   M001 M002 M003 M004 M005 M006 M009 M013 M014 M015 M023 M024 M026 M027
2   M001 M002 M003 M006 M013 M014 M015 M016
3   M003 M013 M014 M015 M016
4   M003 M014 M015 M016
5   M003 M015 M016
6   M003 M013 M014 M015
7   M003 M013 M014 M015 M016
8   M003 M013 M014 M015 M016
9   M001 M002 M004 M005 M006 M007 M009
10  M001 M002 M003 M005 M006 M009 M010 M011 M013 M014 M015 M016
11  M002 M003 M008 M009 M010 M011 M013 M014 M015
12  M001 M002 M003 M004 M005 M006 M009 M013 M014 M015 M016
13  M003 M014 M015 M016
14  M001 M002 M003 M004 M006 M010 M011 M012 M013 M014 M015
15  M003 M014 M015 M016
16  M001 M002 M003 M004 M005 M006 M009 M013 M014 M015 M016 M018 M020 M023 M024
17  M001 M002 M003 M004 M005 M006 M007 M008 M009 M010 M011 M013 M014 M015 M016
18  M003 M013 M014 M015 M016
19  M003 M015 M016
20  M003 M013 M014 M015 M016
21  M003 M014 M015 M016
22  M003 M014 M015 M016
23  M003 M015 M016
24  M003 M014 M015 M016
25  M003 M015 M016
26  M003 M014 M015 M016
27  M003 M013 M014 M015 M016
28  M003 M015 M016
29  M002 M003 M009 M013 M014 M015 M016
30  M003 M015 M016
31  M001 M002 M003 M006 M009 M013 M014 M015 M018 M021 M022
32  M001 M002 M004 M005 M006 M009 M013 M018 M020 M022 M023 M024 M026 M031
33  M002 M003 M004 M005 M006 M008 M009 M010 M011 M013 M014 M015
34  M003 M015 M016
35  M003 M015 M016
36  M001 M003 M005 M013 M014 M015 M016
37  M001 M002 M003 M004 M005 M006 M009 M010 M011 M013 M014 M015 M016
38  M001 M002 M003 M004 M005 M006 M007 M009 M010 M011 M013 M014 M015 M016
39  M002 M003 M007 M009 M010 M011 M012 M013 M014 M015 M016
40  M002 M003 M009 M012 M013 M014 M015 M016
41  M002 M003 M009 M010 M011 M013 M014 M015 M016
42  M002 M003 M013 M014 M015 M016
43  M003 M013 M014 M015 M016
44  M003 M013 M014 M016
45  M013 M014 M015
46  M013 M014 M015 M016
```

Figure 21. All sensor-ID lists between 6 pm and 7 pm of the entire eavesdropping period for deducing kitchen sensors in Tulum.

To deduce kitchen/dining room sensors of Tulum, PDS extracted all indoor activities that occurred between 6 pm and 7 pm from each of the 12 days. The total corresponding sensor-ID lists are 46, which are presented in Figure 21. After applying ARL with the same minSupport on these lists, set {M003, M014, M015, M016} is returned and consequently is deduced as sensors in the kitchen/dining room. To deduce more sensors in the same place, PDS recursively checked the global sensor topology and discovered that M002, M009, M010, M011, M012, and M013 all satisfy the condition. Therefore, they were also deduced as kitchen/dining room sensors. Apparently, the above deduction is correct according to the layout illustrated in Figure

15. However, PDS failed to deduce M007 and M008 since these two sensors do not have sufficient bidirectional edges with the other deduced kitchen/dining room sensors.

Finally, PDS deduced entrance sensors in Tulum by extracting all leave-back activities from the same time period. Four sensors (i.e., M001, M002, M008, and M010) were found in these activities. Since M002 and M010 have already been deduced as kitchen/dining room sensors, they cannot be entrance sensors at the same time. In other words, only M001 and M008 were deduced as entrance sensors. According to Figure 15, it might be incorrect that M001 is an entrance sensor because M005 is more close to the entrance. However, the truth is that the sensing coverage of M001 is large, which even covers the sensing area of M005. Therefore, even though M005 is the one near the entrance, M001 is still the first triggered sensor when the couple living in Tulum enters the house. By further referring to the global sensor topology, PDS knows that these two sensors do not connect to each other, implying that they are two different entrances in Tulum.

Figures 22 and 23, respectively, show the accumulated number of bedroom activities and kitchen/dining room activities in Tulum during the eavesdropping period. We can observe the following routines: (1) Many activities were found in the bedrooms from 08:00 to 09:00 and from 23:00 to 01:00, meaning that the couple living in Tulum may wake up in the first period and go to bed in the second period. (2) All activities in Tulum's bedrooms are crossing-sensor indoor activities, implying that the couple does not have many still activities in their bedrooms. (3) No activities were discovered in the bedrooms from 16:00 to 19:00 but increasing activities were found in the kitchen/dining room during this period, meaning that the couple usually stays in the kitchen/dining room in this period. In fact, Figure 23 even shows the increasing number of activities in the kitchen/dining room between 15:00 and 21:00. In fact, the dining room is also the living room in Tulum (see Figure 15), which explains why the couple is very active in this place during this time period.

Table 2 lists all leave-back activities in Tulum during the eavesdropping period. Similarly to the old woman living in Milan, the couple living in Tulum did not go out together for more than one hour very often since only three leave-back activities were found by PDS. Particularly, we can see that they do not go out before 08:30 and after 17:30. This information might be valuable for malicious attackers. Based on all above deductions for Tulum, we confirm again that PDS is able to expose lots of privacy about this house and the couple.

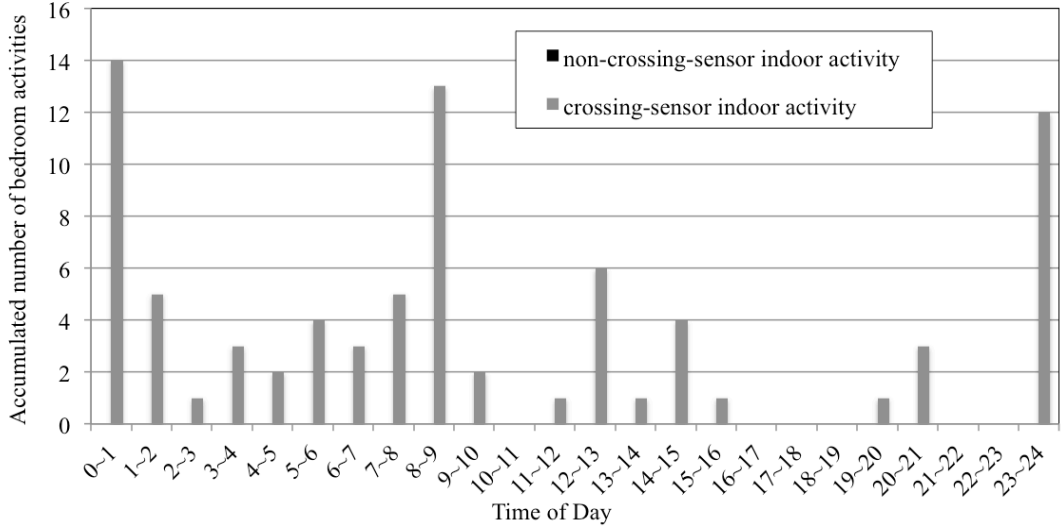

Figure 22. The total number of indoor activities in Tulum's bedrooms during the entire eavesdropping period.

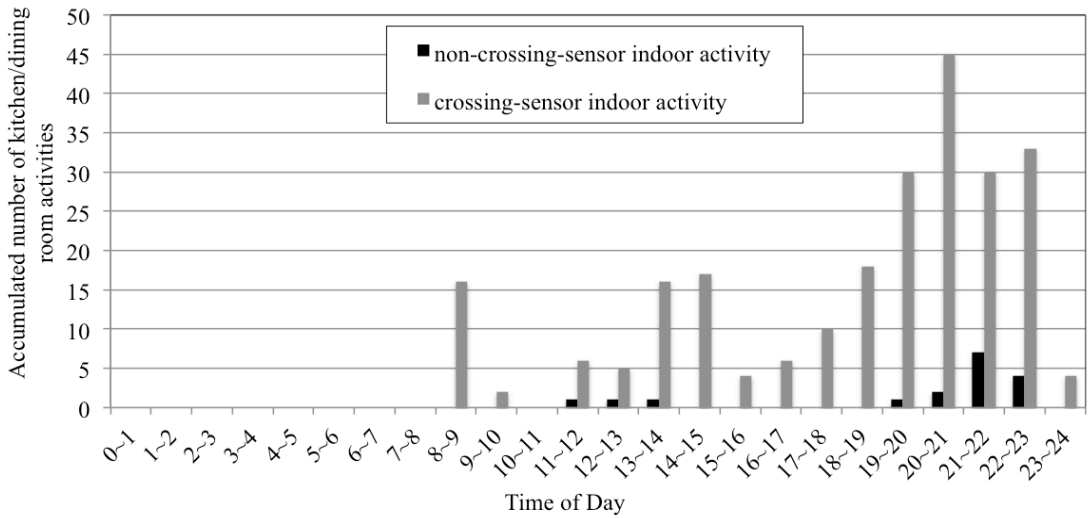

Figure 23. The accumulated number of indoor activities in Tulum's kitchen/dining room during the entire eavesdropping period.

Table II. All leave-back activities in Tulum during the entire eavesdropping period.

| Date (y/m/d) | Time period (24hr) | Sensor ID |
|---|---|---|
| 2009/09/30 | 08:38 to 11:09 | M008 |
| 2009/10/01 | 08:35 to 11:04 | M008 |
| 2009/10/01 | 17:13 to 18:39 | M001 |

## 5. Discussion

In the previous section, we utilize $x = 40$ sec, $y = 10$ sec, and $z = 3600$ sec to identify indoor activities in both Milan and Tulum. In fact, according to the research shown in [36], it is difficult to determine an appropriate time to recognize a human activity because the time span of a human activity could be various. Since PDS utilizes elders' movements to derive a global sensor topology and deduce sensor locations, $x$ cannot be too short. Otherwise, PDS might not be able to capture a whole movement activity. Due to the fact that ARL is utilized by PDS to mine associated sensors in a smart home, the better results can be derived if more identified indoor activities contain at least three different sensors. Based on our experiences, setting $x$ to 40 sec allows PDS to achieve the above identification, and this explains why $x$ is 40 sec in our experiments.

It is not easy to determine a good value for parameter $y$. If $y$ is too small (e.g., 2 sec), PDS is unable to identify many indoor activities, which in turn hampers the deduction performance of PDS. Besides, if $y$ is too small, PDS might not be able to identify a complete movement performed by an elder. On the other hand, if $y$ is too big (e.g., 30 sec), multiple residents' movements might therefore be wrongly identified as an indoor activity by PDS, in turn affecting the deduction of PDS. In this paper, we found that $y = 10$ sec is a reasonable setting. In our future work, we will investigate how to appropriately determine a value for $y$. For parameter $z$, we do not recommend to set a too small value because it is unable for PDS to effectively deduce entrance sensors. However, from the viewpoint of an attacker, he/she can comprehensively analyze elders' leave-back patterns by setting a value depending on his/her preference.

Currently there are several privacy preservation approaches designed for smart homes. The first one is to assemble a robot to periodically trigger a motion and open/close sensor [23] so as to create an illusion for attackers that homeowners are at home. However, this approach does not affect PDS regardless of the trigger frequency. If the robot frequently triggers the sensor, a lot of identified indoor activities will contain this sensor, causing that this sensor has directed edges with almost all the other sensors in the resulting global sensor topology. For a normal sensor deployed in a fixed place, this is impossible because people are unable to move directly from this sensor to each of the rest sensors in a smart home. On the other hand, if the robot infrequently but still periodically triggers the sensor, the directed edges between this sensor and the other sensors will be mostly eliminated by Rule 1 and Rule 2.

Another advance approach is employing a complex robot, suggested by [23], that can move from room to room in a smart home. Clearly this approach successfully creates the illusion that homeowners are at home, but it unfortunately helps PDS to deduce the global sensor topology even faster.

Applying some sophisticated cryptography algorithms (e.g., VPN) to encrypt the network traffics between smart sensors and the smart hub might be the most promising and effective solution to prevent the deduction of PDS since attackers are unable to analyze the corresponding traffic. However, this approach is expensive and might quickly consume energy for smart devices.

## 6. Conclusion and future work

In this paper, we have proposed PDS for deducing global sensor topology, sensor locations, and elders' daily routines from a smart home from an attacker's perspective. According to the experimental results, PDS is capable to infer most bedroom sensors, kitchen/dining room sensors, and entrance sensors. The key factor is that the deduction logic of PDS is based on the two reasonable assumptions, i.e., most people stay in their bedrooms and kitchens/dining rooms during 2 am and 6 am and during 6 pm and 7 pm, respectively. With these two assumptions in mind, PDS can deduce sensors in bedrooms and kitchen/dining room of a smart home. Another key point enabling PDS to deduce more related sensors is the sensor spatial relationships offered by the derived global topology in which most connections between sensors caused by multiple elders' concurrent movements are eliminated by Rules 1 and 2. Employing PDS, elders' privacy could be seriously disclosed without physical invasion. Attackers are able to deduce when and how often elders stay in their bedrooms and kitchens/dining rooms, when elders leave home and come home, etc.

In the future, we would like to further extend PDS with other methods such as deep learning [45] or hidden Markov model (HMM for short) [46] to infer more sensor locations, such as living rooms and toilets, that PDS is currently unable to deduce. In addition, we would like to find out the best values for setting parameter $y$.

**Acknowledge**

This work was partially supported by the project IoTSec – Security in IoT for Smart Grids, with number 248113/O70 part of the IKTPLUSS program funded by the Norwegian Research Council.

**References**

[1] F.K., Santoso and N.C. Vun, "Securing IoT for smart home system," In *2015 IEEE International Symposium on Consumer Electronics* (*ISCE*), *IEEE*, 2015, June, pp. 1–2.

[2] How the Internet of Things will affect security & privacy. http://www.businessinsider.com/internet-of-things-security-privacy-2016-8?r=US&IR=T&IR=T, 2019 (accessed 27 February 2019).

[3] D.J. Cook and S.K. Das, "Smart environments: Technology, protocols and applications," vol. 43, *John Wiley & Sons*, 2004.

[4] J. Lundström, E. Järpe, and A. Verikas, "Detecting and exploring deviating behaviour of smart home residents," *Expert Systems With Applications*, vol. 55, 2016, pp. 429–440.

[5] P. Rashidi, D.J. Cook, L.B. Holder, and M. Schmitter-Edgecombe, "Discovering Activities to Recognize and Track in a Smart Environment," *IEEE transactions on knowledge and data engineering*, vol. 23, no. 4, 2011, pp. 527–539.

[6] L. Lu, Q.l. Cai, and Y.J. Zhan, "Activity Recognition in Smart Homes," *Multimedia Tools and Applications*, 2016, pp. 1–18.

[7] M.A.A. Pedrasa, T.D. Spooner, and I.F. MacGrill, "Coordinated scheduling of residential distributed energy resources to optimize smart home energy services," *IEEE Transactions on Smart Grid*, vol. 1, no. 2, 2010, pp. 134–143.

[8] H. Park, C. Basaran, T. Park, and S. H. Son, "Energy-Efficient Privacy Protection for Smart Home Environments Using Behavioral Semantics," *Sensors*, vol. 14, no. 9, 2014, pp. 16235–16257.

[9] A.M.M. Ali, N.M. Ahmad, and A.H.M. Amin, "Cloudlet-based cyber foraging framework for distributed video surveillance provisioning," In *2014 Fourth World Congress on Information and Communication Technologies* (*WICT*), *IEEE*, 2014, December, pp. 199–204.

[10] K. Fleming, P. Waweru, M. Wambua, E. Ondula, and L. Samuel, "Toward quantified small-scale farms in africa," *IEEE Internet Computing* vol. 20, no.3, 2016, pp. 63–67

[11] M. Ryu, J. Yun, T. Miao, I.Y. Ahn, S.C. Choi, and J. Kim, "Design and implementation of a connected farm for smart farming system," In *SENSORS, 2015 IEEE*, pp. 1–4.

[12] H. Kumarage, I. Khalil, A. Alabdulatif, Z. Tari, and X. Yi, "Secure data analytics for cloud-integrated internet of things applications", *IEEE Cloud Computing*, vol. 3, no. 2, 2016, pp. 46–56.

[13] A. Sajid, H. Abbas, and K. Saleem, "Cloud-assisted IoT-based SCADA systems security: A review of the state of the art and future challenges," *IEEE Access*, vol. 4, 2016, pp. 1375–1384.

[14] W. Trappe, R. Howard, and R.S. Moore, "Low-energy security: Limits and opportunities in the internet of things," *IEEE Security & Privacy*, vol. 13, no.1, 2015, pp. 14–21.

[15] H. Chae, J. Park, H. Song, Y. Kim, and H. Jeong, "The IoT based automate landing system of a drone for the round-the-clock surveillance solution," In *2015 IEEE International Conference on Advanced Intelligent Mechatronics* (*AIM*), *IEEE*, 2015, July, pp. 1575–1580.

[16] Z. Liu and T. Yan, "Study on multi-view video based on IOT and its application in intelligent security system," In *Proceedings 2013 International Conference on Mechatronic Sciences, Electric Engineering and Computer* (*MEC*)*, IEEE*, 2013, December, pp. 1437–1440.

[17] M. Taneja, "A framework to support real-time applications over ieee802.15.4 DSME," In *2015 IEEE Tenth International Conference on Intelligent Sensors, Sensor Networks and Information Processing* (*ISSNIP*), *IEEE*, 2015, April, pp. 1–6.

[18] P. Bellavista, G. Cardone, A. Corradi, and L. Foschini, "Convergence of MANET and WSN in IoT urban scenarios," *IEEE Sensors Journal*, vol. 13, no. 10, 2013, pp. 3558–3567.

[19] P. Rashidi and D.J. Cook, "COM: A Method for Mining and Monitoring Human Activity Patterns in Home-based Health Monitoring System," *ACM Transactions on Intelligent Systems and Technology*, vol. 4, no. 4, 2013, pp. 64.

[20] M. Mather, L. A. Jacobsen, and K. M. Pollard, "Aging in the United States," *Population Bulletin*, vol. 70, no. 2, 2015.

[21] UN Department of Economic and Social Affairs. Population Division. World population in 2300. New York, United Nations; 2004. https://warwick.ac.uk/fac/soc/pais/research/researchcentres/csgr/green/foresight/demography/united_nations_world_population_to_2300.pdf, 2019 (accessed 27 February 2019).

[22] D. J. Cook, "Learning setting-generalized activity models for smart spaces," *IEEE Intelligent Systems*, vol. 27, issue: 1, 2012, pp. 32–38.

[23] M. R. Schurgot, D. A. Shinberg, and L. G. Greenwald, "Experiments with security and privacy in IoT networks," *2015 IEEE 16th International Symposium on a World of Wireless, Mobile and Multimedia Networks (WoWMoM),* 2015, pp. 1–6.

[24] K. Yoshigoe, W. Dai, M. Abramson, and A. Jacobs, "Overcoming invasion of privacy in smart

[25] home environment with synthetic packet injection," In *TRON Symposium (TRONSHOW),* IEEE, 2015, pp. 1–7.
[25] Hacking The Z-Wave protocol with a HackRF. http://www.rtl-sdr.com/hacking-the-z-wave-protocol-with-a-hackrf/ , 2019 (accessed 27 February 2019).
[26] Z-Wave Sniffer. http://www.suphammer.net/zwave_sniffer, 2019 (accessed 27 February 2019).
[27] Pretty Much Every Smart Home Device You Can Think of Has Been Hacked, https://slate.com/technology/2014/12/the-internet-of-things-is-a-long-way-from-being-secure.html, 2019 (accessed 27 February 2019).
[28] A. Cui and S.J. Stolfo, "A quantitative analysis of the insecurity of embedded network devices: results of a wide-area scan," In *Proceedings of the 26th Annual Computer Security Applications Conference*, ACM, 2010, December, pp. 97–106.
[29] L. Chen, J. Hoey, C.D. Nugent, D.J. Cook, and Z. Yu, "Sensor-Based Activity Recognition," *IEEE Transactions on Systems, Man, and Cybernetics, Part C* (*Applications and Reviews*), vol. 42, no.6, 2012, pp. 790–808.
[30] N.C. Krishnan and D.J. Cook, "Activity recognition on streaming sensor data," *Pervasive and mobile computing*, vol. 10, 2014, pp. 138–154.
[31] P. Lukowicz, J. Ward, H. Junker, M. St¨ager, G. Tr¨oster, A. Atrash, and T. Starner, "Recognizing workshop activity using body worn microphones and accelerometers," In Proceedings of the Second International Conference on Pervasive Computing (Pervasive), pages 18–32, Vienna, Austria, April 2004.
[32] J. Lester, T. Choudhury, N. Kern, G. Borriello, and B. Hannaford, "A hybrid discriminative/generative approach for modeling human activities," In *Proceedings of the Nineteenth International Joint Conference on Articial Intelligence* (IJCAI), pp. 766–772, Edinburgh, Scotland, July–August 2005.
[33] U. Maurer, A. Smailagic, D.P. Siewiorek, and M. Deisher, "Activity recognition and monitoring using multiple sensors on different body positions," In *International Workshop on Wearable and Implantable Body Sensor Networks*, IEEE, 2006, April, pp. 113–116.
[34] L. Zhao, X. Wang, G. Sukthankar, and R. Sukthankar, "Motif discovery and feature selection for crf-based activity recognition," In *2010 20th International Conference on Pattern Recognition* (*ICPR*)*,* IEEE, 2010, August, pp. 3826–3829,
[35] T. Inomata, F. Naya, N. Kuwahara, F. Hattori, and K. Kogure, "Activity recognition from interactions with objects using dynamic bayesian network," In *Proceedings of the 3rd ACM International Workshop on Context-Awareness for Self-Managing Systems*, ACM, 2009, May, pp. 39–42.
[36] T. van Kasteren and B. Kröse, "Bayesian activity recognition in residence for elders," *Proceedings of the IET International Conference on Intelligent Environments*, 2007, pp. 209–212
[37] T. Gu, Z. Wu, X. Tao, H. K. Pung, and J. Lu, "epsicar: An emerging patterns based approach to sequential, interleaved and concurrent activity recognition," In *IEEE International Conference on Pervasive Computing and Communications, PerCom 2009*, IEEE, 2009, March, pp. 1–9.
[38] T. Gu, S. Chen, X. Tao, and J. Lu, "An unsupervised approach to activity recognition and segmentation based on object-use fingerprints," *Data & Knowledge Engineering*, vol. 69, no.6, pp. 533–544.
[39] P. Rashidi, and D. J. Cook, "Mining sensor streams for discovering human activity patterns over time," In *2010 IEEE 10th International Conference on Data Mining* (*ICDM*), IEEE, 2010, December, pp. 431–440.
[40] C. Giannella, J. Han, J. Pei, X. Yan, and P. S. Yu, "Mining frequent patterns in data streams at multiple time granularities," *MIT Press*, 2003, Ch. 3.
[41] We Are Creatures of Habit. https://www.psychologytoday.com/blog/creatures-habit/200907/we-are-creatures-habit, 2019 (accessed 27 February 2019).
[42] K. Lai, and C. Narciso, "Support vs. confidence in association rule algorithms," *Proceedings of the OPTIMA Conference, Curicó*, 2001.
[43] WSU CASAS Datasets. http://ailab.wsu.edu/casas/datasets/, 2019 (accessed 27 February 2019).
[44] D. Cook, and M. Schmitter-Edgecombe, "Assessing the quality of activities in a smart environment," *Methods of information in medicine*, vol. 48, no. 5, 2009, pp. 480–485.
[45] Deep Learning for Java. https://deeplearning4j.org/index.html, 2019 (accessed 27 February 2019).
[46] Hidden Markov model. "https://en.wikipedia.org/wiki/Hidden_Markov_model", 2019 (accessed 27 February 2019).
[47] M.C. Lee, J.C. Lin, and O. Owe, "Privacy Mining from IoT-Based Smart Homes," In *International Conference on Broadband and Wireless Computing, Communication and Applications*, Springer, Cham, 2018, October, pp. 304-315.